\documentclass[%
 reprint,
 superscriptaddress,
 graphicx,
 amsmath,amssymb,
 aps,prr,
 longbibliography,
]{revtex4-2}

\usepackage{graphicx}% Include figure files
\usepackage{dcolumn}% Align table columns on decimal point
\usepackage{bm}% bold math
\usepackage[breaklinks=true,colorlinks=true,linkcolor=blue,urlcolor=blue,citecolor=blue]{hyperref}
\usepackage{upgreek}

\begin{document}

%\preprint{APS/123-QED}

\title{Data-driven discovery of reduced plasma physics models from fully-kinetic simulations}%\thanks{A footnote to the article title}%
%alternative titles
% Generation of high-quality ion beams from radiation pressure acceleration of thin targets
% High-quality radiation pressure ion acceleration from thin targets via the mitigation of electron heating
\author{E. P. Alves}
\email[]{epalves@physics.ucla.edu} 
\affiliation{%
 High Energy Density Science Division, SLAC National Accelerator Laboratory, Menlo Park, CA 94025, USA
}% 
\affiliation{Department of Physics and Astronomy, University of California, Los Angeles, CA 90095, USA}
\author{F. Fiuza}
\email[]{fiuza@slac.stanford.edu}
\affiliation{%
 High Energy Density Science Division, SLAC National Accelerator Laboratory, Menlo Park, CA 94025, USA
}

%\title{Data-driven discovery of reduced plasma physics models from fully-kinetic simulations}% Force line breaks with \\
% \thanks{A footnote to the article title}%

%\author{E. P. Alves}
%\email{epalves@physics.ucla.edu}
%\affiliation{%
% High Energy Density Science Division, SLAC National Accelerator Laboratory, Menlo Park, CA 94025, USA
%}
%\affiliation{%
%Department of Physics and Astronomy, University of California, Los Angeles, CA 90095, USA
%}

%\author{F. Fiuza}%
%\email{fiuza@slac.stanford.edu}
%\affiliation{%
% High Energy Density Science Division, SLAC National Accelerator Laboratory, Menlo Park, CA 94025, USA
%}%

%\date{\today}% It is always \today, today,
             %  but any date may be explicitly specified

\begin{abstract}
At the core of some of the most important problems in plasma physics --- from controlled nuclear fusion to the acceleration of cosmic rays --- is the challenge to describe nonlinear, multi-scale plasma dynamics. The development of reduced plasma models that balance between accuracy and complexity is critical to advancing theoretical comprehension and enabling holistic computational descriptions of these problems. Here we report the data-driven discovery of accurate reduced plasma models, in the form of partial differential equations, directly from first-principles particle-in-cell simulations. We achieve this by using an integral formulation of sparsity-based model-discovery techniques and show that this is crucial to robustly identify the governing equations in the presence of discrete particle noise. We demonstrate the potential of this approach by recovering the fundamental hierarchy of plasma physics models --- from the Vlasov equation to magnetohydrodynamics. Our findings show that this data-driven methodology offers a promising new route to accelerate the development of reduced theoretical models of complex nonlinear plasma phenomena and to design computationally efficient algorithms for multi-scale plasma simulations.
\end{abstract}

%\keywords{Suggested keywords}%Use showkeys class option if keyword
                              %display desired
\maketitle

%\tableofcontents

\section{\label{sec:intro}Introduction}

Plasmas --- hot, ionized gases of electrons and ions that make up most of the observable Universe --- exhibit rich many-body dynamics that span a vast range of scales. It is widely recognized that kinetic processes occurring at microscopic scales can strongly influence and control plasma phenomena at large (system size) scales. Notable examples include the role of microphysical instabilities on the deterioration of plasma confinement in nuclear fusion devices \cite{Goeler74,Kadomtsev75} and the role of microphysical turbulence in controlling the acceleration and propagation of energetic cosmic rays in astrophysical environments \cite{Blandford87,Bell04}. The holistic understanding of these problems remains a long-standing scientific challenge; addressing it requires a better theoretical description of the interplay between the different processes and the ability to model the resulting nonlinear plasma dynamics across the different scales. Fully-kinetic simulations \cite{Dawson1983,Birdsall1991} can provide first-principles descriptions of the plasma dynamics, but at tremendous computational cost and complexity that prohibits modeling the full range of scales for most systems of interest. On the other hand, fluid simulations are commonly used to capture the large-scale plasma behavior, but miss the important microphysical processes. The development of reduced models that capture the \emph{essence} of the interplay between the microscopic kinetic processes and large-scale fluid behavior is therefore key to enabling multi-scale plasma modeling for a variety of applications.

Despite significant efforts and important progress in the last decades, the multitude of concurrent physical processes and their inherently nonlinear character has limited the theoretical development of reduced plasma models. Indeed, the majority of existing models are based on asymptotic limits \cite{Zocco2011,Crestetto2020} or on linear approximations \cite{hammetperkins1990} of the reduced physics, both of which limit their range of validity and often break down (locally or intermittently) in many problems of interest. Fully-kinetic simulations play an important role in the study of nonlinear kinetic plasma phenomena, but it remains unclear how to distill the insights captured from the data of such simulations into practical theoretical models.

Data-driven techniques from the field of machine learning are offering powerful new ways of building models of nonlinear dynamical systems from data that can greatly complement more traditional theoretical approaches. Specifically, symbolic- \cite{Bongard2007,Schmidt2009,Udrescu2020,Cranmer2020} and sparse-regression (SR) \cite{Wang2011,Brunton2016,Rudy2017,Schaeffer2017a} techniques have been identified as promising routes for inferring \emph{interpretable} and \emph{generalizable} nonlinear differential equations (both ordinary and partial differential equations, ODEs and PDEs) directly from time-series data. These techniques seek parsimonious models that balance between accuracy and complexity, providing insight into the underlying physics and enabling a direct connection with analytic theory. SR in particular has been shown to efficiently handle high-dimensional and multi-variate data of dynamical systems \cite{Brunton2016, Rudy2017}, making it potentially well suited for plasma dynamics \cite{Dam2017, Kaptanoglu2021}. However, the application of these techniques in plasma physics remains largely unexplored. In particular, an important issue that is still unclear is whether SR can be used to infer accurate reduced descriptions of nonlinear kinetic plasma processes --- essential for improving the aforementioned multi-scale plasma models ---  from first-principles fully-kinetic simulations. As shown in previous works \cite{Brunton2016,Rudy2017}, even modest ($\sim 1\%$) levels of (artificially added) noise can corrupt the identification of the underlying dynamical equations, which can be problematic when using inherently noisy data from particle-based kinetic simulations.

Here we show that SR is a viable approach for the data-driven development of interpretable reduced plasma models from first principles, fully-kinetic particle-in-cell (PIC) simulations. We illustrate the potential of SR through the recovery of the fundamental hierarchy of plasma physics models --- from the kinetic Vlasov equation to single-fluid magnetohydrodynamics (MHD) --- based solely on spatial and temporal data of nonlinear plasma dynamics from PIC simulations.
In order to robustly handle the significant noise levels in particle-based data, we reformulate the sparsity-based model discovery methodology \cite{Brunton2016,Rudy2017,Schaeffer2017a} to identify the underlying PDEs in their integral form; we show that this is crucial, particularly to capture low-frequency plasma phenomena that is embedded in high-frequency discrete particle noise for which the differential formulation is found to fail. We further show that the hierarchy of Pareto-optimal models naturally obtained by this approach provides insight into the dominant physical processes underlying the plasma dynamics, which can guide the development of tailored reduced models for a given application.

\section{\label{sec:results}Results}
% \subsection{\label{sec:level2}Results}

We examine how SR can be used to discover reduced PDE models of plasma dynamics from the data of first-principles PIC simulations. The PIC method provides a self-consistent and fully kinetic (particle-based) description of a plasma (see Appendix \ref{app:pic}), and hence it is a good starting point from which reduced models of collisionless or weakly collisional plasmas can be constructed.

\begin{figure*}[t!]
\begin{center}
\includegraphics[width=\textwidth]{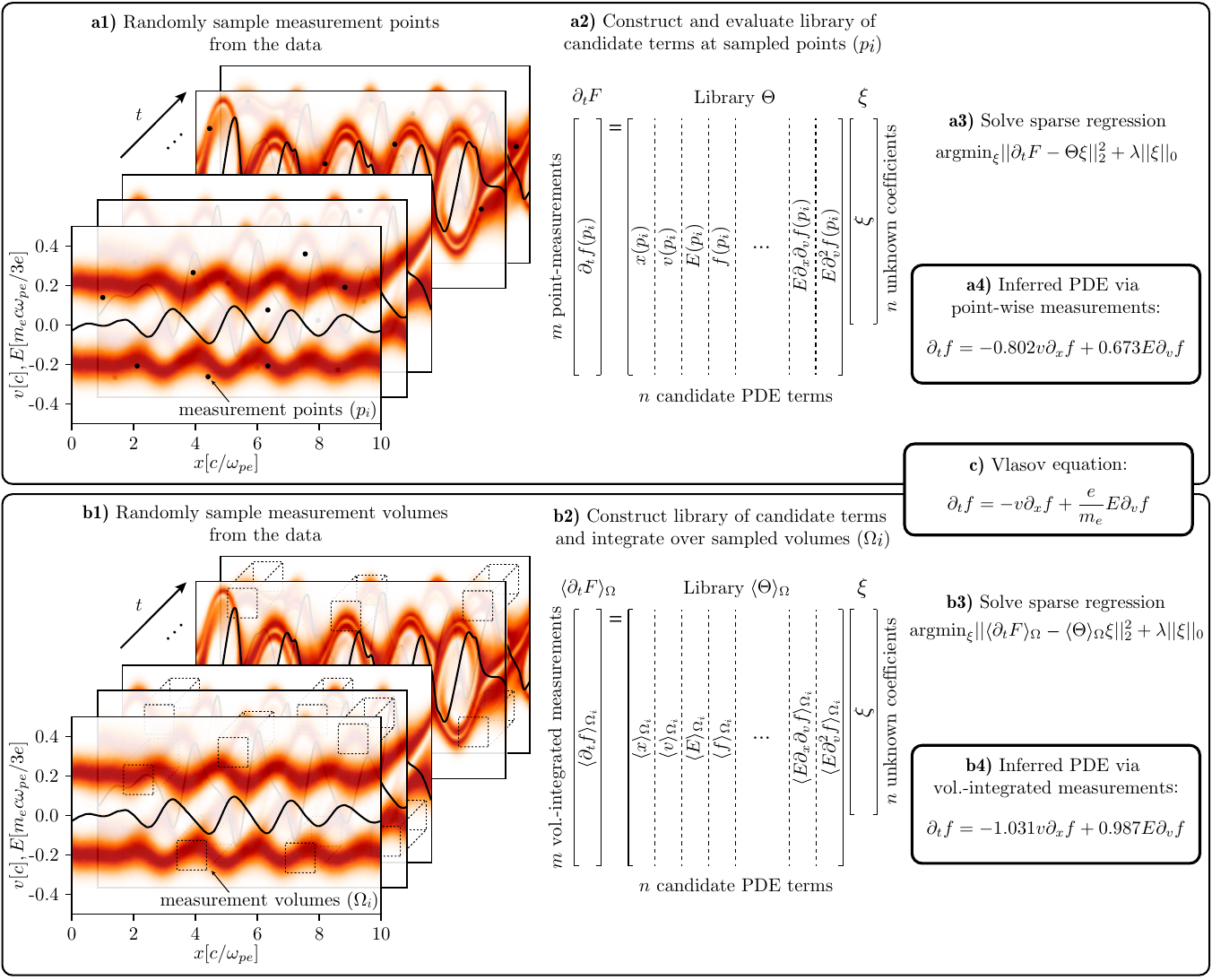}
\caption{
\textbf{Inferring the Vlasov equation from PIC simulation data using SR}. The data consists of temporal snapshots of the plasma distribution function in phase space ($f$, represented by the red color density plots), and the self-consistent plasma electric field ($E$, represented by the black solid curves), undergoing the two-stream instability; the discrete particle noise in the distribution function data is estimated to be $\sqrt{\mathrm{VAR}(f - \tilde{f})/\mathrm{VAR}(\tilde{f})}\simeq 4\%$, where $\mathrm{VAR}$ denotes variance and $\tilde{f}$ is the Gaussian-filtered distribution function with $\sigma=1$. The inference procedure consists of (1) sampling the data at random locations, (2) evaluating the library of candidate PDE terms at sampled locations, and (3) solving a SR problem to select the most parsimonious combination of PDE terms that best describes the data (in a least-squares sense). The inferred PDEs are shown in (4) and can be compared to the true Vlasov equation in (c); note that the simulated data used in this example corresponds to counter-streaming electron populations which have $e/m_e = 1$ in the normalized units of the PIC simulation. Two different strategies for collecting measurements from data are presented in the top and bottom rows: (a) point-wise measurements of each candidate term $Q$ [as used in \cite{Rudy2017}], and (b) volume-integrated measurements of each candidate term $\langle Q\rangle_{\Omega_i} = \int_{\Omega_i}\mathrm{d}\Omega~Q$. The latter effectively corresponds to the identification of the underlying PDE in its integral form [rather than its differential form (a)].
}
\label{fig1}
\end{center}
\end{figure*}

\textbf{The Vlasov equation.}
We begin by considering the problem of recovering one of the most fundamental equations in plasma physics -- the kinetic Vlasov equation [Figure~\ref{fig1}(c)] -- which describes the evolution of the distribution function $f(\mathbf{x},\mathbf{v},t)$ of a collisionless plasma in phase space (where $\mathbf{x}$, $\mathbf{v}$ and $t$ are respectively space, velocity and time coordinates). We analyze the plasma distribution function (constructed from the simulated particles) and associated electromagnetic fields of a system of counter-propagating electron streams undergoing the electrostatic two-stream instability \cite{ONeil71} [Figure \ref{fig1} (a1); see Appendix \ref{app:datagen} for a detailed description of simulation parameters]. This data is representative of prototypical nonlinear plasma dynamics in phase space. Based on this data, we aim to infer the PDE that governs the evolution of the distribution function $\partial_t f$ (where $\partial_\alpha$ denotes partial differentiation with respect to coordinate $\alpha$). As in Refs. \cite{Brunton2016,Rudy2017,Schaeffer2017a}, this inference is posed as a SR problem, where we seek to find the sparsest PDE (with least number of terms) that best describes the data (in a least squares sense) from a large space of candidate PDEs. 

We proceed by constructing the space of possible PDEs, which is represented by a large library ($\Theta$) of candidate PDE terms. The choice of candidate terms is guided by fundamental physical symmetries and domain knowledge of the system under study. In the present example, for instance, possible terms may include the distribution function itself and its gradients up to some prespecified order $d$ ($f$, $\partial_x f$, $\partial_v f$, ..., $\partial_x^d f$, $\partial_v^d f$), the electric field and its gradients ($E$, $\partial_x E$, ..., $\partial_x^d E$), and the phase space coordinates ($x$ and $v$). We may also construct candidate nonlinear terms by taking polynomial combinations of the previous terms up to order $p$. Such nonlinear terms give rise, for instance, to advective (e.g., $v\partial_x f$) and electromagnetic pressure gradient (e.g., $E\partial_xE \propto \partial_xE^2$) terms; note that nonpolynomial nonlinearities (e.g. trigonometric, exponential and logarithmic functions) can also be included if prior knowledge or intuition suggest that these may be important candidate terms \cite{Brunton2016}. We denote the total number of candidate terms by $n$. In this particular case, we consider both derivatives and polynomial nonlinearities up to second order ($d=2$ and $p=2$), yielding a total of $n=66$ candidate PDE terms; we note that we have also tested using larger libraries containing higher-order polynomial nonlinearities (up to $d=4$, yielding a total of $n=1001$ terms) and obtained similar results to those described below.

In Figure \ref{fig1} (a), we outline the algorithmic procedure proposed in \cite{Rudy2017}, known as PDE-FIND (the PDE generalization of SINDy \cite{Brunton2016}), where each of the $n$ candidate terms in the library and $\partial_t f$ are evaluated on a subset of $m$ point-wise locations in the data [$p_i = (t_i, v_i, x_i)$, with index $i$ denoting the $i\mathrm{th}$ point], that randomly sample the dynamics in phase space and in time; note that neighboring points are utilized to estimate derivative terms at each point $p_i$. We utilize second-order centered-finite-differences to evaluate derivative terms, and we do not filter or smooth the data. We sample the data at $m=1.25\times10^5$ random points ($\sim 1\%$ of the total generated data). We then infer the underlying PDE by solving the SR problem in Figure~\ref{fig1} (a3), where we seek the sparsest vector of coefficients $\xi$ that minimizes $||F_t - \Theta \xi||^2_2$. Specifically, we compute sparse solutions for $\xi$ using a variation of the sequential thresholded least-squares algorithm proposed in \cite{Brunton2016} (see Appendix \ref{app:sralgorithm}), and use $10$-fold cross-validation to determine the optimal level of sparsity that balances between model accuracy and complexity.

This procedure leads to the successful identification of the correct terms in the Vlasov equation, which correspond to the advection of the distribution function in space ($v\partial_xf$) and in momentum ($E\partial_v f$) [Figure~\ref{fig1}(a4)]. The inferred coefficients of these terms, however, present larger errors ($\sim20-30\%$). Their theoretical values are $-1$ and $1$ (in the normalized simulation units of the data), but are inferred to be $-0.8025$ and $0.6726$, respectively.

We find that the main reason for the large errors in the inferred PDE coefficients is due to poor estimation of derivative terms on noisy data; straightforward finite differencing is ill-conditioned, as the noise is amplified upon differentiation \cite{Chartrand2011}. Indeed, for this reason, even modest levels of noise ($\sim 1\%$) have been shown to corrupt the PDE identification procedure, making it a key challenge for this methodology \cite{Brunton2016,Rudy2017}. The PIC simulation data considered here inevitably contains noise due to discrete particle effects, i.e. fluctuations associated with the finite number of simulated particles [$\propto 1/\sqrt{N_{ppc}}$, where $N_{ppc}$ is the number of particles per (spatial) cell; in the example of Figure~\ref{fig1} we used $N_{ppc} = 10^4$ which is typical for these simulations]. Note that while the number of simulated particles may be sufficient to properly capture the physics of interest (e.g. the growth and nonlinear evolution of the two-stream instability), it may be insufficient to reduce the noise to tolerable levels for successful sparse identification of the underlying PDE. It is therefore crucial to develop effective techniques that permit the robust inference of PDEs from noisy particle-based data without increasing the $N_{ppc}$ used in these simulations, which is computationally very costly.

Some of the techniques considered by previous works consisted in denoising the data (e.g. via Gaussian smoothing) prior to the identification process, or using regularized numerical differentiation techniques (e.g. polynomial differentiation) \cite{Brunton2016,Rudy2017,Schaeffer2017a}. We find these methods to have limited success when handling data from PIC simulations, as summarized in Table~\ref{tab1}. The main difficulty lies in controlling the bias-variance trade-off of these methods, which if inadequate can corrupt the identification process entirely. This is illustrated by the results of Gaussian smoothing on the identification of the Vlasov equation in Table~\ref{tab1}. Indeed, while moderate smoothing ($\sigma=1$, which corresponds to using a Gaussian kernel with standard deviation equal to the resolution of the data) was found to reduce the error of the inferred Vlasov coefficients from 25\% error to 6\%, the error rapidly rose again to 16\% with a slightly increased smoothing level. Differentiating a locally-fitted low-degree polynomial to the data, as used in \cite{Rudy2017}, is also found to be ineffective on PIC simulation data.

\begin{table}[ht]
\centering
 \begin{tabular}{l c } 
 \hline
                                     & Avg. coeff. error \\
 \hline
 CFD (no Gaussian filter)            & 25\%         \\
CFD + Gaussian filter ($\sigma = 1$) & 6\%          \\
CFD + Gaussian filter ($\sigma = 2$) & 16\%         \\
PI, $3\mathrm{rd}$-degree, $m = 5$   & 37\%         \\
PI $3\mathrm{rd}$-degree, $m = 7$    & 60\%         \\
\textbf{Integral formulation}        & \textbf{2\%} \\
 \hline
 \end{tabular}
\caption{\textbf{Impact of noise-mitigation strategies on error of inferred coefficients of the Vlasov equation.} The average coefficient errors for the Vlasov equation (inferred from the same data shown in Figure\ref{fig1}, which has an estimated noise level of $\sim 4\%$) are presented for the following strategies: centered finite differences (CFD) on data with varying levels of Gaussian filtering (no filter, $\sigma=1$ and $\sigma=2$; note that Gaussian filtering is applied in both phase space coordinates and in time); polynomial interpolation (PI) where derivatives are computed by differentiating a polynomial of degree $3$ that is locally fitted to the data over a range of $m>4$ points (we present results for $m=5$ and $m=7$); and using the integral formulation strategy, with $\Omega_i = 5\Delta t\times 5\Delta v\times 5\Delta x$.}
\label{tab1}
\end{table}

\textbf{Sparse identification of PDEs in their integral form.}
In order to more effectively overcome the challenges posed by data noise on model discovery, we reformulate the problem as that of identifying the underlying PDE in its integral form. The motivation being that numerical integration compensates the noise amplification induced by differentiation. The use of an integral formulation (or weak formulation) for the inference of governing ODEs from time series data has been previously proposed in Refs. \cite{Crutchfield1987,Schaeffer2017b}. During the development of this work, other groups independently recognized the potential for extending this integral formulation to identification of PDEs \cite{Reinbold2020, Messenger2021b}. The integral formulation strategy that we independently explored is illustrated in
Figure~\ref{fig1} (b). Each candidate term in the PDE (including the time derivative term) is now evaluated using centered-finite-differencing and then numerically integrated over compact volumes on the data $\Omega_i = \{ (x,v,t): |x-x_i|<w_x/2 \wedge |v-v_i|<w_v/2 \wedge |t-t_i|<w_t/2 \}$, where $w_\alpha$ is the length of the edges of $\Omega_i$ along the $\alpha$ coordinate. Thus, each measurement on the data corresponds to the integration of each candidate term over randomly distributed volumes $\Omega_i$. In Figure~\ref{fig1}(b), we adopt this strategy and randomly sample $10^3$ cubic volumes of $5\Delta x\times5\Delta v\times5\Delta t$ in phase space and time, corresponding to the same total of $m=1.25\times10^5$ points used for the point-wise strategy in Figure~\ref{fig1}(a); here we choose the volume size to be much larger than a single cell while remaining smaller than the characteristic scale of variation in the data. The same library of candidate PDE terms used earlier is evaluated and integrated over these volumes and the SR problem in Figure~\ref{fig1}(b3) is solved. Interestingly, this strategy leads to the correct identification of the Vlasov PDE to within $\sim2\%$ error on the inferred coefficients, significantly outperforming the original differential formulation strategy (with point-wise evaluation of the terms on the data), even when data smoothing or regularized differentiation techniques are used (Table~\ref{tab1}). The inferred coefficient errors can be further reduced by increasing the size of the integration volumes. Overall, we find that the integral formulation always leads to much reduced errors when compared with the standard differential formulation, even when using a very large number of particles to produce cleaner data. A detailed comparison between the noise sensitivity properties of the integral and standard differential formulations as a function of the number of simulated particles and size of integration volumes is provided in Appendix \ref{app:impactNppc}.
%on data produced using  (using more particles).
%A detailed discussion on how the intrinsic data noise and the size of integration volumes impact the performance of the integral formulation, as well as a comparison with the standard differential formulation on data produced using  (using more particles).
%as discussed in the Supplemental Material.
We note that the integral formulation approach presented here corresponds to a special case of the more general weak formulation proposed in \cite{Reinbold2020,Messenger2021b}. These works integrate over the product between each candidate PDE term and an arbitrary test function with compact support, while we have used a simple ``box''-function (constant and non-zero within a compact domain and zero outside of it). The use of more general test functions that smoothly vanish at the boundaries of the domain can further reduce the contributions of numerical differentiation error at the boundaries of the integration volumes. Nevertheless, our results demonstrate the effectiveness of the integral formulation strategy on mitigating numerical differentiation errors without introducing deleterious bias in the data. This is a crucial advantage for a robust sparse PDE identification based on noisy data, such as that from partiscle-based simulations.

\textbf{Signatures of successful PDE identification.}
It is important to emphasize that successful PDE identification using SR relies on the appropriate construction of the library of candidate terms $\Theta$. In the previous example, domain knowledge was used to guide the selection of the candidate basis terms in which the representation of the dynamics becomes parsimonious. In addition, the library of candidate terms was complete, meaning that it contained all the terms necessary to fully describe the dynamics in the data. When applying this approach to less well understood problems, however, it may not always be clear what the correct choice of basis terms is, or if the library is complete. It is therefore important to understand the empirical signatures that indicate successful or unsuccessful sparse identification of the underlying PDE.

\begin{figure}[t!]
\begin{center}
\includegraphics[width=\linewidth]{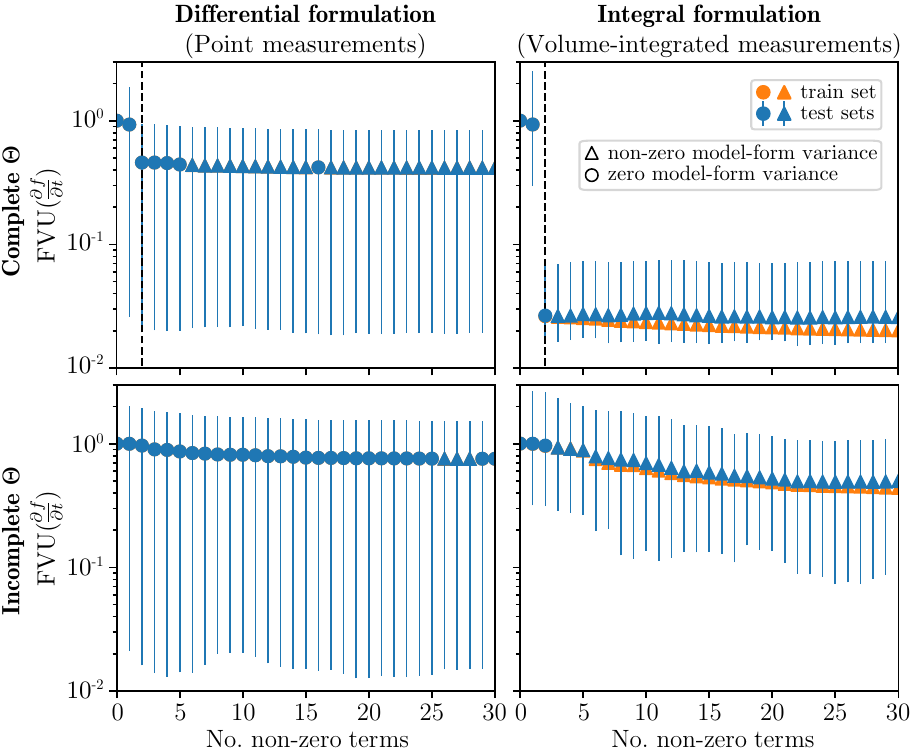}
\caption{\textbf{Signatures of successful/unsuccessful PDE identification}. Pareto analysis of model accuracy [measured by the Fraction of Variance Unexplained (FVU)] versus complexity (measured by the number of non-zero terms) obtained using the differential formulation strategy (integral formulation strategy) on the left (right) column, and for the case of a complete (incomplete) library of terms on the top (bottom) row. The FVU is given by the ratio between the model's mean squared error and the variance of the dependent variable ($\partial f/\partial t$). Orange (blue) markers represent model evaluations on the training (test) data; error bars represent the minimum and maximum test errors encountered during cross-validation. Circle (triangle) markers represent zero (non-zero) model-form variance (i.e. variance in the obtained sparsity pattern) obtained during cross-validation. The vertical dash lines in the top row mark the optimal trade-off between accuracy and complexity, and correspond to identification of the correct two terms of the Vlasov equation.
}
\label{fig2}
\end{center}
\end{figure}

These signatures can be observed in the behavior of the model accuracy/complexity curve that is traced by the SR procedure. Using the identification of the Vlasov equation as an example, Figure~\ref{fig2} illustrates the accuracy/complexity curves that represent successful identification (top row), obtained by using a complete library $\Theta$, and the cases of unsuccessful identification of the Vlasov equation (bottom row) due to the explicit removal of an important dynamical term (spatial advection $v\partial_x f$) from $\Theta$, representing an incomplete library scenario; the original differential and new integral formulations are shown on the left and right columns, respectively, for comparison. Note that successful PDE identification is characterized by a pronounced inflection in the curve, where the model error suddenly rises due to the thresholding of an important dynamical term, marking the optimal trade-off between model accuracy and complexity. It is interesting to further note that the integral formulation reveals a far more pronounced inflection in the curve, and a much steeper rise in error at the optimal accuracy/complexity trade-off [Figure ~\ref{fig2} (b)] compared to the original differential formulation [Figure ~\ref{fig2} (a)]. This is due to the improved evaluation of the PDE terms in the presence of noise, and further highlights the advantage of the integral formulation from the point of view of facilitating and improving the overall robustness of the PDE identification procedure; see Figure~\ref{figS3} and further discussion in Appendix~\ref{app:impactNppc} about the impact of varying the size of the integration volumes on the resulting accuracy/complexity curves. In the absence of an \emph{important} dynamical term, the model accuracy/complexity curve no longer displays a clear inflection region [Figures~\ref{fig2} (c,d)]. Instead, the model error is found to steadily increase as we progressively decrease the complexity (increase the sparsity) of the model. In addition, high variance in the model form (i.e. the sparsity pattern) and in the identified model coefficients is also observed during cross-validation. All these indicators point to poor or unreliable PDE identification, and suggest that the library of candidate terms must be reexamined, reformulated or simply expanded (e.g. higher degree of nonlinearity and/or order of differentiation).

\begin{figure*}[t!]
\begin{center}
\includegraphics[width=\textwidth]{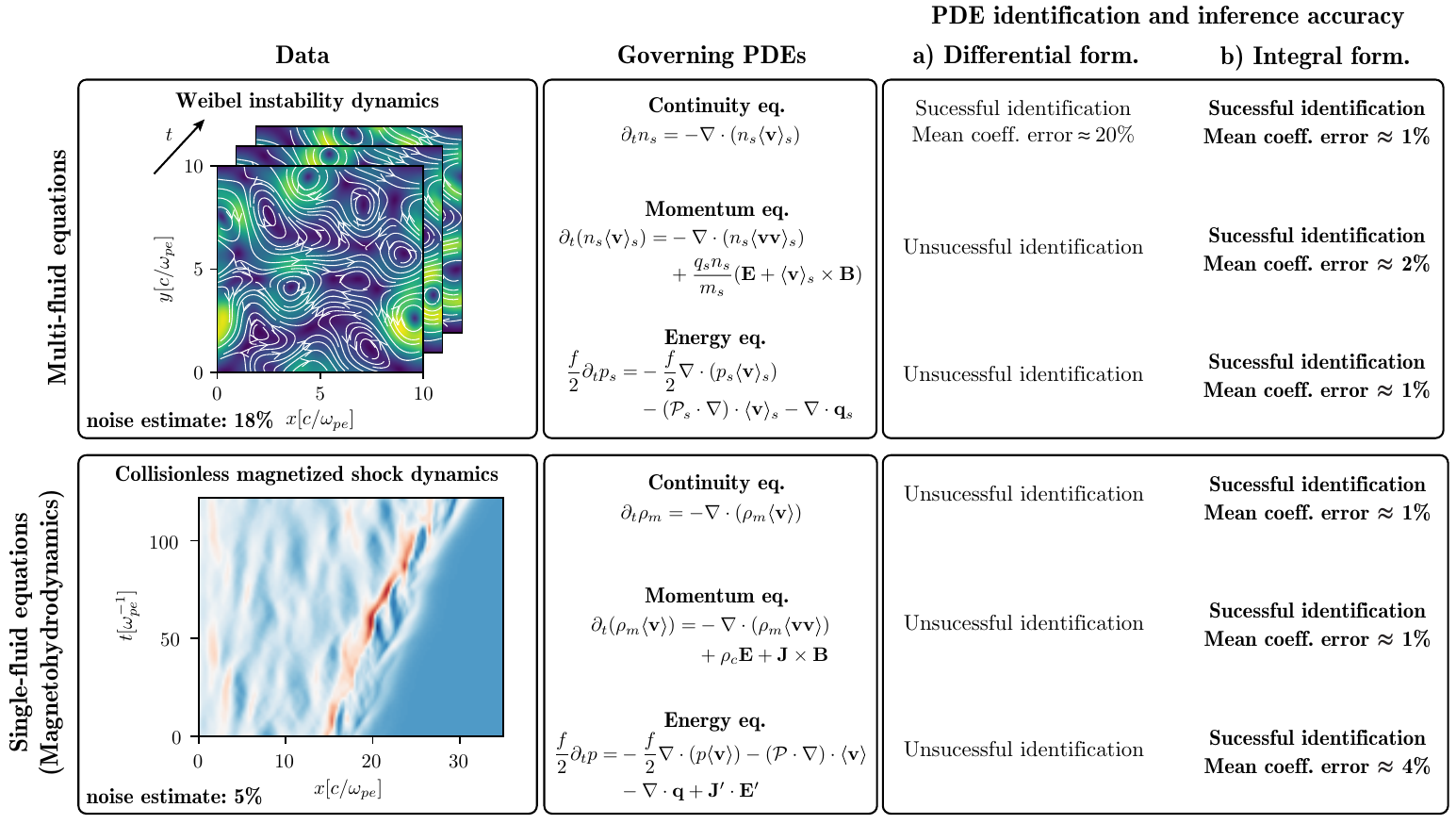}
\caption{\textbf{Inferring the multi-fluid and single-fluid (MHD) equations from PIC simulation data.} Data of the development of the Weibel instability and of the propagation of a collisionless magnetized shock are used to infer the multi-fluid and single-fluid equations, respectively; the noise levels are estimated to be $\sqrt{\mathrm{VAR}(n_e - \tilde{n_e})/\mathrm{VAR}(\tilde{n_e})}\simeq 18\%$ for the Weibel data and $\sqrt{\mathrm{VAR}(\rho_m - \tilde{\rho_m})/\mathrm{VAR}(\tilde{\rho_m})}\simeq 5\%$ for the shock data, where $n_e$ and $\rho_m$ are the electron number density and average mass density of the of the plasma, respectively. The differential formulation strategy (using straightforward centered finite differences to evaluate derivative terms) is shown to be incapable of correctly identifying the underlying equations (with the exception of the multi-fluid continuity equation, albeit with significant errors in the coefficients). The integral formulation strategy, on the other hand, is shown to robustly recover the correct PDEs and with high (percent level) accuracy.
}
\label{fig3}
\end{center}
\end{figure*}

\textbf{The multi-fluid and single-fluid plasma equations.}
In order to explore the potential of this approach to infer progressively more reduced plasma descriptions, we aim to recover the well-known plasma fluid equations directly from PIC simulation data. These equations describe the evolution of the velocity-moments of the plasma distribution function (the first three moments correspond to the mass, momentum and energy densities of the fluid). They can be written for individual plasma species as coupled fluids (the so-called multi-fluid equations), or they can be made to describe the average fluid behaviour of all plasma species (the so-called single-fluid equations, or MHD). The fluid equations can be derived from the kinetic Vlasov equations and form an infinite hierarchy of exact coupled conservation equations for each fluid moment.
In practice, this infinite hierarchy is truncated after the first few moments by imposing an \emph{approximate} closure relation --- i.e. a relation that expresses the evolution of the highest-order moment considered in terms of lower-order moments. It is the level at which this hierarchy of coupled moment equations is truncated and the assumptions that underpin the closure relation that control the physical approximations of plasma fluid descriptions. Indeed, commonly used closures allow the description of the plasma dynamics near local thermodynamic equilibrium and at large spatial and temporal scales, but miss kinetic plasma processes occurring at microscopic scales.
 
It remains a long-standing challenge to understand how to extend the fluid equations to encapsulate microscopic plasma effects in a computationally efficient manner, which would ultimately enable multi-scale modeling of problems ranging from fusion to astrophysical plasmas. The SR methodology explored here offers a new approach to discovering such extensions to the fluid equations by inferring them directly from the data of first-principles PIC simulations. In the following, we show how SR can be used to infer conservation equations for both multi- and single- fluid moments, and guide the development of improved kinetic-fluid closure models.

\begin{figure*}[t!]
\begin{center}
\includegraphics[width=0.8\textwidth]{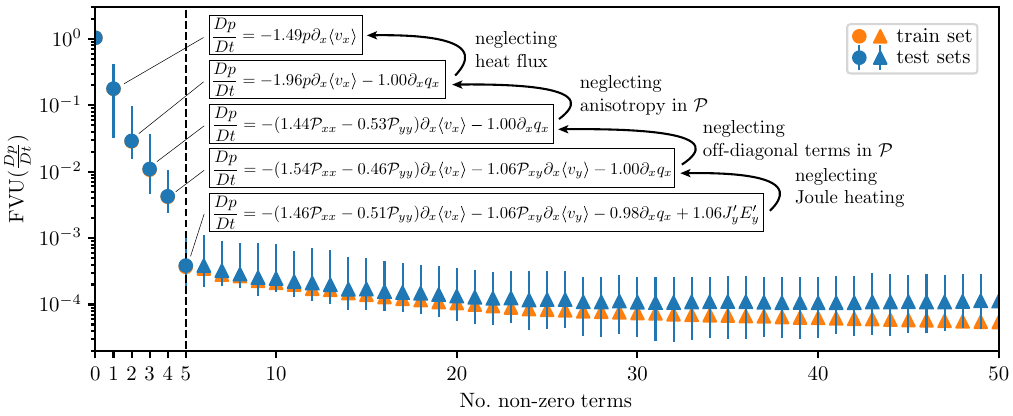}
\caption{\textbf{Hierarchy of Pareto-optimal models obtained for the MHD energy density equation from the magnetized shock data}. Pareto analysis of the FVU of each model [$\mathrm{MSE(model)}/\mathrm{VAR}(Dp/Dt)$] versus  the number of non-zero terms. The markers and error bars have the same meaning as in Figure~\ref{fig2}. 10-fold cross-validation indicates that the optimal accuracy/complexity trade-off is obtained for a model of $5$ terms (marked by the vertical dashed line), which corresponds to the recovery of the full form of the MHD energy density equation. A hierarchy of reduced models is obtained at lower model complexities, with the arrows indicating the physical meaning of the successive approximations identified by the SR procedure.
}
\label{fig4}
\end{center}
\end{figure*}

The inference of both multi-fluid and single-fluid plasma equations (specifically for the first three moments) from PIC simulation data is shown in Figure~\ref{fig3}. The multi-fluid equations are inferred from data of the development and nonlinear evolution of the electromagnetic Weibel instability \cite{Weibel59,Fried59} (Figures~\ref{fig3} top). This is a fundamental plasma instability associated with the anisotropy of the plasma velocity distribution function, and plays an important role in magnetic field amplification in both astrophysical \cite{Medvedev99} and laboratory \cite{Silva2002,Fiuza2020} environments. As for the single-fluid (MHD) equations, we use data of the formation and propagation of a collisionless magnetized shock \cite{Forslund1971} (Figure \ref{fig3} bottom); details about the simulations parameters, data sampling and design of the library of candidate terms used for the SR can be found in Appendices \ref{app:datagen}, \ref{app:sampling} and \ref{app:thetadesign}. These examples push the SR PDE identification methodology to far more challenging regimes, involving electromagnetic phenomena and a larger set of dynamical variables. In addition, the magnetized shock propagation data is predominantly characterized by slow, low-frequency plasma dynamics, implying that derivatives/gradients of the data are small, and hence are highly susceptible to being corrupted by the intrinsic PIC noise. Indeed, for these reasons, we observe that the differential formulation is unsuccessful in identifying the correct form of the fluid equations, as summarized in Figure~\ref{fig3}; poor results are still obtained when using denoising and regularized numerical differentiation techniques.
% (see Supplemental Information).
The integral formulation, however, overcomes these challenges and robustly recovers the correct form of the fluid equations and with high accuracy ($\sim 1\%$ error in inferred coefficients). This highlights the potential of the integral formulation to effectively capture slow and large scale dynamics in the presence of high-frequency phenomena, which is essential to build coarse-grained models from data.

As previously noted, the fluid equations shown in Figure~\ref{fig3} are exact conservation laws for the first three moments of the distribution function, but they do not represent a closed system of equations --- the energy equation depends on the heat flux ($\mathbf{q}$), a higher order moment of the distribution function. We have verified (not shown here) that we can proceed to infer the PDE governing the evolution of the heat flux given data of higher order moments, and we could in principle continue to infer this hierarchical system of equations to arbitrary order. In practice, however, one aims to find an approximate closure relation to truncate this infinite hierarchy to lowest moment order possible (low model complexity) while still being sufficiently accurate to describe the dynamics of interest.

\textbf{Closure of the fluid equations.}
To illustrate how SR can be used to infer fluid closure models from fully kinetic simulations, we focus on the previous example of the formation and propagation of a collisionless magnetized shock.

Collisionless magnetized shocks are ubiquitous in space and astrophysical environments and represent a prototypical example of a large-scale macroscopic plasma phenomenon where microphysical kinetic processes play an important role in their dynamics. In recent years, kinetic simulations of shocks have been playing a critical role in elucidating the range of microphysical processes that control dissipation of the directed flow kinetic energy into randomized thermal energy, magnetic field amplification, and acceleration of nonthermal particle populations \cite{Silva2003, Spitkovsky2008, Sironi2010,Caprioli2014}. Unfortunately, kinetic simulations are limited to spatial and temporal scales that are vastly smaller than the large scale dynamics of shocks in space and astrophysical systems. For this reason, simulations based on reduced plasma descriptions are key to modeling the dynamics of shocks in large scale systems. In this context, MHD is often used as the framework of choice \cite{GiacaloneJokipii2007, Samsonov2007, Skillman2013}. In scenarios where particle acceleration is important, MHD can be complemented by the use of test particles \cite{Kang&Jones1997, Beresnyak2011} or by recently developed MHD-PIC frameworks \cite{Bai2015, vanMarle2018}, where the feedback of energetic particles (described by PIC) on the background fluid (described by MHD) is captured on large-scales. However, in all these cases, the description of the background fluid relies on oversimplified closures that cannot capture the impact of the dominant microscopic effects on the shock evolution. Here, we show that SR offers a powerful new approach to leverage the detailed dynamical data from first-principles kinetic simulations to inform the dominant microscopic processes and guide the development of improved kinetic-fluid closures that encapsulate their effects in MHD.

This is exemplified in Figure~\ref{fig4}, which shows the model accuracy/complexity curve obtained during the inference of the MHD energy density equation (using the integral formulation) from the magnetized shock data. The pronounced inflection of this curve at a model complexity of $5$ terms marks the optimal trade-off between complexity and accuracy, and corresponds to the complete conservative form of the MHD energy density equation (with mean coefficient error of $\simeq4\%$). At lower model complexities, a hierarchy of reduced models is obtained that reflects successive approximations to the MHD energy density equation that are directly informed by the data. Each of these approximations has a clear physical meaning as indicated in Figure~\ref{fig4}.
In this particular example, we find that the most dynamically important terms are (in ascending order) Joule heating, gyro-viscous effects (associated with off-diagonal elements of the pressure tensor), pressure anisotropy, heat flux, and (isotropic) compressional heating. The SR approach thus provides important insight into the dominant physical processes and allows us to quantify the error associated with neglecting each model term, which are critical to guide the development of tailored reduced models for any given application.

It is interesting to observe that among the hierarchy of inferred Pareto-optimal models, the widely used adiabatic closure approximation is automatically identified as the simplest (yet least accurate) model for the MHD energy density equation for the plasma dynamics in this data. The adiabatic closure corresponds to the $1-$term model in Figure~\ref{fig4}, which contains only the compressional heating term ($p\partial_x\langle v_x\rangle$). This approximate model truncates and closes the hierarchy of fluid equations by neglecting the gradient of the heat flux ($\partial_x q_x$) term, but misses $\simeq20\%$ of the variance in the total time-derivative of the plasma energy density.
% (see Appendix \ref{app:spatiotemporal} for a discussion on the spatiotemporal error distribution of the adiabatic closure model and other hierarchical models).
Note that the coefficient of $1.49$ corresponds to the inferred adiabatic index. If the heat flux were truly negligible, however, the adiabatic index would be expected to take on the value of $2$ since the shock heating in this system occurs only in two velocity degrees of freedom (heating along the direction of the background magnetic field does not occur in this geometry). The reduced value of the inferred adiabatic index is thus a consequence of the heat flux not being entirely negligible in this scenario, and results from the SR procedure attempting to compensate its absence so as to best describe the data.

In addition to analyzing the overall accuracy of a given model (as shown in the accuracy/complexity curves), it is instructive to examine the spatiotemporal distribution of the model error on the data to gain a deeper understanding of its ability to describe the physics/dynamics in the data and also to diagnose missing physics in the model (model bias). Such analysis reveals, for example, that the pressure anisotropy and Joule heating terms are important primarily in the shock transition region and can be neglected outside this region. A detailed discussion is provided in Appendix~\ref{app:spatiotemporal} (and Figure~\ref{figS4}). The spatiotemporal model error distribution can thus be valuable in guiding the choice of appropriate approximations when developing reduced models.

Having identified the dominant physical processes in the energy density equation and the importance of heat flux in the dynamics of the magnetized shock, the stage is set for the development of more accurate closure models;
the same SR approach can be used to uncover accurate approximations of the heat flux and higher-order fluid moments as nonlinear functions of the lower-order moments, which will be the focus of future work.
In this context, it is worth discussing some of the key aspects for the development of improved closure models that should be considered in future studies.
First, and as previously discussed, a central challenge lies in choosing an appropriate basis of nonlinear candidate terms for the library $\Theta$ that can efficiently approximate the underlying closure. In general, the nonlinear functional form of the underlying closure relation may be non-polynomial in nature, making the choice of a polynomial basis of candidate terms for $\Theta$ inappropriate or incomplete. Domain knowledge may often suggest specific non-polynomial functional forms that can be incorporated in the library. For instance, rational function nonlinearities \cite{Mangan2016} (as found in the well-known CGL closure \cite{CGL1956}), or non-local (integral) terms (as in linear Landau-fluid closures \cite{hammetperkins1990}) can increase the descriptive capacity of the library to better approximate the dynamics of interest. Second, the explicit incorporation of physical constraints that reflect fundamental physical symmetries in the SR procedure can be important to ensure physical consistency and generalizability of the inferred closures. This has recently been demonstrated in the context of low-dimensional reduced order models for hydrodynamics \cite{Loiseau2018} and MHD \cite{Kaptanoglu2021}. Finally, care must be given to the choice of training data. In addition to preparing an ensemble of simulations that are representative of the dynamics and physical regime of interest, it is also important to ensure that the inferred reduced models are not biased by the limited domain sizes or scale separation of the kinetic simulations. It is thus important to verify that by progressively increasing the simulation domain size, there is asymptotic convergence of the inferred closure model.

\section{Discussion}
We have shown that SR is a viable approach for extracting \emph{interpretable} and \emph{generalizable} reduced models of complex plasma dynamics from the data of first-principles kinetic simulations. This data-driven methodology can accelerate theoretical insight into out-of-equilibrium and highly nonlinear plasma dynamics (e.g. the nonlinear evolution of instabilities \cite{Davidson72,Katsouleas88,Quest1996,Schekochihin2008}), which so far have challenged traditional analytical approaches. The \emph{interpretable} form of these data-driven PDEs will facilitate the connection between the identified terms and basic physical processes, and will naturally stimulate theoretical efforts to ``reverse engineer'' these models starting from lower-level frameworks.

We further envision that this SR methodology can be used to develop computationally efficient reduced models for multi-scale plasma simulations, which remains a grand challenge in computational plasma physics. Indeed, the hierarchy of progressively simpler Pareto-optimal models produced by this methodology provides a powerful tool to determine the optimal trade-off between model accuracy and complexity for a specific application. Important examples include the development of improved fluid closures that encapsulate desired kinetic effects and subgrid models of coarse-grained phenomena, such as anomalous resistivity and transport. These are essential ingredients for the development of more accurate multi-scale algorithms for applications that range from whole fusion device modeling to global simulations of space and astrophysical systems.

It is noteworthy that the integral formulation strategy, which was shown to be crucial for the robust inference of PDEs from noisy PIC simulation data, can be applied more broadly to the data of other particle-based simulation techniques. These include molecular dynamics, direct simulation Monte Carlo, and other N-body simulations, commonly used in areas that range from atomic physics to cosmology. Indeed, even within the context of plasma physics, we note that while this work focused on extracting reduced descriptions of collisionless/weakly collisional plasma dynamics from the data of kinetic PIC simulations, other particle-based simulation methods such as molecular dynamics \cite{Marciante2017} and extended PIC-Monte Carlo \cite{TakizukaAbe1977,Alves2021} can be more appropriate to obtain reduced models of collisional and strongly-coupled plasmas using the same data-driven methodology. Moreover, while we have focused here on simulated data, it is an exciting prospect to explore the application of this methodology to experimental laboratory and spacecraft data. The remarkable progress in plasma diagnostics is enabling spatially and temporally resolved measurements with unprecedented quality \cite{Glenzer2016,Gekelman2016,Lindqvist2016,Fox2016}, creating opportunities to leverage this methodology in the near future.

In summary, we have presented a data-driven framework capable of distilling interpretable plasma physics models from the increasingly abundant and complex data of plasma dynamics. Our results open a new avenue to accelerate theoretical developments of nonlinear plasma phenomena, and to leverage these insights to design computationally efficient algorithms for multi-scale plasma simulations.

\begin{acknowledgments}
The authors would like to thank S. Brunton, N. Kutz, W. Dorland and N. Loureiro for helpful discussions. The authors acknowledge the OSIRIS Consortium, consisting of UCLA and IST (Portugal) for the use of the OSIRIS 4.0 framework. Simulations were run on Mira (ALCF) and Cori (NERSC) through ALCC awards. This work was supported by the U.S. Department of Energy SLAC Contract No. DE-AC02-76SF00515, by the U.S. DOE Early Career Research Program under FWP 100331, and by the National Science Foundation Grants No. NSF PHY-2108087 and No. NSF PHY-2108089.
\end{acknowledgments}

\appendix

\section{\label{app:pic}Particle-in-cell (PIC) method.}
The PIC method is a particle-based simulation technique that aims to capture the kinetic microphysics of plasmas. Formally, the PIC method \cite{Dawson1983,Birdsall1991} solves the Klimontovich equation \cite{Klimontovich1967} (for finite size particles) coupled to Maxwell's equations. The numerical procedure consists of solving Maxwell's equations on a spatial grid using the current and charge densities that are obtained by weighting discrete plasma particles onto the grid; the particles are then advanced via the Lorentz force associated with their self-consistent collective electric and magnetic fields. Thus, to the extent that quantum mechanical effects can be neglected, the PIC method provides a first principles description of plasma dynamics.

\section{\label{app:datagen}Data generation.}
All PIC simulation data used in this work were obtained using OSIRIS 4.0 \cite{Fonseca2002a,Fonseca2008}, which is a massively parallel, relativistic, electromagnetic, explicit PIC code. Details of the physical and numerical parameters used to produce the data presented in the main text are given below.

\textit{Two-stream instability.} We simulated two symmetrically counter-streaming flows of electrons in a background of immobile ions in 1D1V (one spatial dimension and 1 velocity degree of freedom). We considered warm electron flows with fluid velocity $\mathbf{v}_0 = \pm 0.2c~\mathbf{\hat{x}}$, thermal velocity of $v_{th} = 0.05c$, and density $n_e = n_0/2$. The immobile ions had number density $n_i = n_0$, providing charge and current neutral initial conditions. The two-stream instability was triggered by thermal fluctuations of the plasma.

The size of the simulated domain was $L = 10c/\omega_{pe}$ (where $c/\omega_{pe}$ is the electron skin depth), spatially and temporally resolved with $\Delta x = 0.039 c/\omega_{pe}$ and $\Delta t = 0.038 \omega_{pe}^{-1}$, respectively. Total simulated time was $T = 50\omega_{pe}^{-1}$, allowing the two-stream instability to fully enter the nonlinear regime. Periodic boundary conditions were used. We used quadratic particle shapes, and performed multiple simulations where we varied the number of particles per cell in the range $10^1 - 10^5$ to investigate the impact of particle noise on the PDE identification procedure. The simulation data used for inference of the Vlasov equation in the main text corresponds to the case of $10^4$ particles per cell. The inference of the Vlasov equation for varying number of particles per cell is presented in Appendix \ref{app:impactNppc}.

From these simulations we collected snapshots of the phase space data of the plasma distribution function ($f$) and the spatial distribution of the plasma electric field ($E$). Each snapshot of the phase space data captured a domain corresponding to $[0~c/\omega_{pe}, 10~c/\omega_{pe}]\times[-0.5c, +0.5c]$, resolved by a $256\times256$ grid; the cell size in velocity space was chosen to resolve both the thermal velocity and the finest structures that develop in the distribution function in the nonlinear phase of the instability. Each snapshot of the electric field distribution over the entire domain was recorded on a 1D grid with $256$ cells. Both the phase space and electric field data were recorded at every other simulation time step, i.e. the temporal resolution of the data was $2\Delta t$. This corresponded to a total of $\simeq 650$ snapshots for each diagnostic.

\textit{Weibel instability.} We simulated two counter-streaming flows of electrons in a background of immobile ions in 2D3V. We considered a warm electron flow with fluid velocity $\mathbf{v}_{0-up} = 0.8c \mathbf{\hat{z}}$ (flowing in the z-direction, perpendicular to the simulation domain), thermal velocity of $v_{th} = 0.15c$, and density $n_{e-up} = n_0/3$, and a counter-propagating electron flow with fluid velocity $\mathbf{v}_{0-down} = -0.4c~\mathbf{\hat{z}}$, thermal velocity of $v_{th} = 0.15c$, and density $n_{e-down} = 2n_0/3$. The background immobile ions had number density $n_i = n_0$, providing charge neutral and current neutral initial conditions. The Weibel instability was triggered by thermal fluctuations of the plasma.

The size of the simulated domain was $10\times10~(c/\omega_{pe})^2]$, spatially resolved with $128\times128$ cells. The simulation time step was $\Delta t = 0.038~\omega_{pe}^{-1}$, and the evolution of the system was simulated up to $T = 30\omega_{pe}^{-1}$, enough time for the Weibel instability to enter the nonlinear regime. Periodic boundary conditions were used. We used cubic particle shapes and $36$ particles per cell per species.

From this simulation we collected the spatiotemporal evolution of the first few moments of the distribution function of each individual electron flow: the number density ($n_s$), momentum density ($n_s\langle \mathbf{v}\rangle_s$), momentum flux density tensor ($n_s\langle \mathbf{vv}\rangle_s$), and energy flux density ($n_s\langle v^2\mathbf{v}\rangle_s$) of each species. We also collected the self-consistent electric ($\mathbf{E}$) and magnetic fields ($\mathbf{B}$). The temporal and spatial resolution of the data used for the inference of the multi-fluid equations was the same as that of the simulation, $\Delta t$ and $\Delta x$.

\textit{Magnetized collisionless shock.} We simulated the formation and propagation of a magnetized collisionless shock in 1D2V. Specifically, we considered a perpendicular shock, where the magnetic field is perpendicular to the shock normal, and considered electron-positron pair plasma. The shock was formed by colliding fresh (upstream) magnetized plasma with a reflective wall at the left boundary of the simulation domain. The upstream plasma had a fluid velocity of $\mathbf{v}_0 = - 0.2c~\mathbf{\hat{x}}$, thermal velocity of $v_{th} = 0.025c$, number density $n_0$, and carried with it a perpendicular magnetic field with $\mathbf{B}_0 = 0.04 m_ec\omega_{pe}/e ~\mathbf{\hat{y}}$.

The size of the simulated domain was $40 ~ c/\omega_{pe}$, resolved with $512$ cells, and the simulation time step was $\Delta t = 0.076~\omega_{pe}^{-1}$. The evolution of the system was simulated up to $T = 350~\omega_{pe}^{-1}$, enough time to capture the cyclic reformation process of the shock front multiple times, and to capture its propagation across a significant fraction of the simulated domain. Fresh upstream magnetized plasma was continuously injected from the right boundary (with open boundary conditions for the fields), while a reflective boundary condition for particles (and conducting for fields) was used for the left boundary of the domain. We used quadratic particle shapes and $10^3$ particles per cell per species.

We collected the spatiotemporal evolution of the first few moments of the distribution function, from the number density up to the energy flux density, of each plasma species (electrons and positrons), as well as the self-consistent electric ($\mathbf{E}$) and magnetic ($\mathbf{B}$) fields. To infer the single-fluid (MHD) equations, we computed the single-fluid (MHD) variables by averaging the moments over the individual species. We thus obtained the single-fluid mass density ($\rho_m \equiv m_e(n_e + n_p)$), momentum density ($\rho_m\langle \mathbf{v}\rangle \equiv m_e(n_e\langle \mathbf{v}\rangle_e + n_p\langle \mathbf{v}\rangle_p)
$), momentum flux density tensor ($\rho_m \langle \mathbf{vv}\rangle \equiv m_e(n_e\langle \mathbf{vv}\rangle_e + n_p\langle \mathbf{vv}\rangle_p$)), and energy flux density ($ \rho_m \langle v^2\mathbf{v}\rangle \equiv m_e(n_e\langle v^2\mathbf{v}\rangle_e + n_p\langle v^2\mathbf{v}\rangle_p)$); we also computed the charge ($\rho_c \equiv e(n_p-n_e)$) and current ($\mathbf{J} \equiv e(n_p\langle\mathbf{v}\rangle_p-n_e\langle\mathbf{v}\rangle_e)$) densities. The temporal resolution of the collected data was $2\Delta t$, while full resolution was used in space ($\Delta x$). 

Note that due to the details of the underlying numerical scheme implemented in OSIRIS, many of the simulation quantities have a relative offset by a half a time step or spatial step. We therefore use linear interpolation to center all variables in time and space prior to the identification procedure.

\section{\label{app:sralgorithm}Sparse regression algorithm.}
In this work we utilize a variation of the sequential thresholded least-squares algorithm proposed in the SINDy framework \cite{Brunton2016} to compute sparse solutions of $\xi$. The original algorithm consisted in solving a least-squares regression for $\xi$ and then setting to zero all coefficients smaller than given threshold $\lambda$; this procedure is repeated on the remaining non-zero coefficients until convergence is reached. The value of the threshold $\lambda$ is progressively increased to obtain increasingly sparse solutions of $\xi$. Our approach consists in starting with an initial least-squares regression on $\xi$ and to eliminate the term with the smallest coefficient norm at each step; this procedure is repeated until the desired level of sparsity is reached. We found that the advantage of this approach is that it allows to systematically sweep across all levels of sparsity in $\xi$, whereas the former method can miss some of the possible solutions (by simultaneously thresholding more than one coefficient at each iteration). As a result, this sparse regression procedure traces cleaner accuracy versus complexity curves, which facilitate the identification of the correct model during cross-validation.

\begin{figure*}[t!]
\begin{center}
\includegraphics[width=1\textwidth]{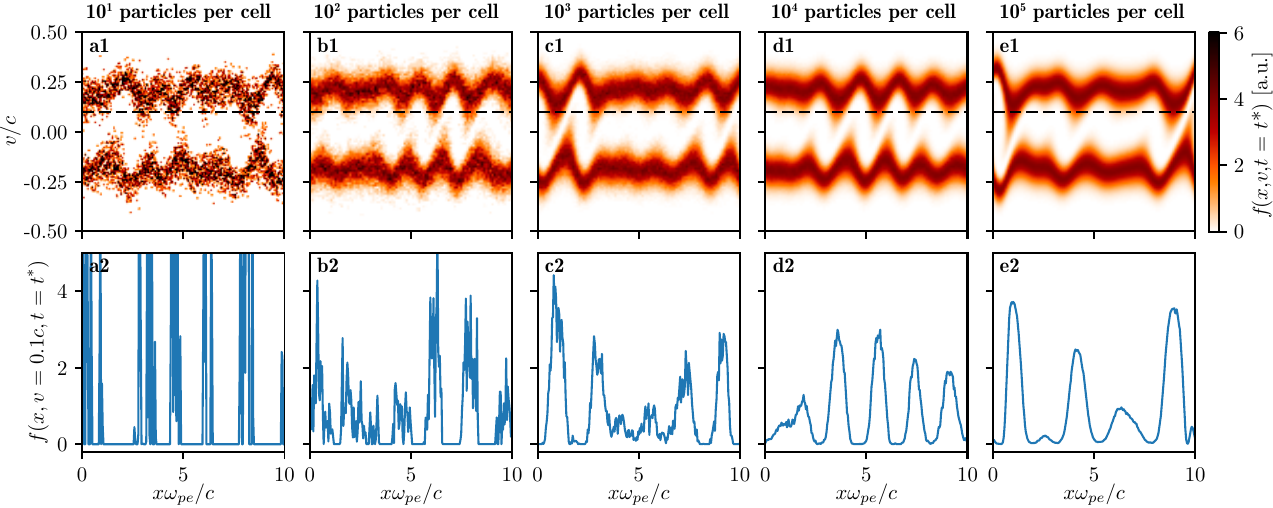}
\caption{\textbf{Visualization of the noise/fluctuation levels in the two-stream instability data for varying numbers of particles per cell.} Figures $a-e$ correspond to the cases of $10^1-10^5$ particles per cell, respectively. The top row displays snapshots of the electron distribution function in phase space around the time of the onset of the nonlinear phase of the instability; note that since the instability grows from thermal noise and the noise decreases with the increasing number of particles per cell, these snapshots are taken at slightly different times, but correspond to similar stages in the development of the instability. The bottom row shows a line-out of the distribution function at constant $v/c = 0.1$ (indicated by the dashed lines in the top row), highlighting how pronounced the noise levels are in the data.
}
\label{figS1}
\end{center}
\end{figure*}

\section{\label{app:impactNppc}Impact of number of particles per cell and size of integration volumes on error of inferred PDE}
The number of particles per cell ($N_{ppc}$) used in PIC simulations controls the amplitude of discrete particle fluctuations in the simulated plasma. While the number of numerical particles used in typical simulations is, in general, can be orders of magnitude lower than the number of physical particles in the real physical systems they aim to describe, the $N_{ppc}$ is chosen so that the numerical fluctuations remain small enough as to not affect the physics/dynamics of interest in the simulation. These fluctuations are illustrated in the Figure~\ref{figS1}, which displays data of the plasma distribution function from two-stream instability simulations with varying $N_{ppc} = 10^1 - 10^5$; the line-outs of the distribution function at constant velocity shown in the bottom row highlight how pronounced these fluctuations are, particularly at low $N_{ppc}$. Note that, in PIC simulations, $N_{ppc}$ refers to the number of particles per spatial cell and not phase space cell; in this cases below the average number of particles per phase space cell (which directly controls the level of fluctuations of the distribution function) is $\sim N_{ppc}/256$, where $256$ is the number of cells used to resolve the phase space velocity axis (see Appendix~\ref{app:datagen} for data generation details).

\begin{figure}[t!]
\begin{center}
\includegraphics[width=\linewidth]{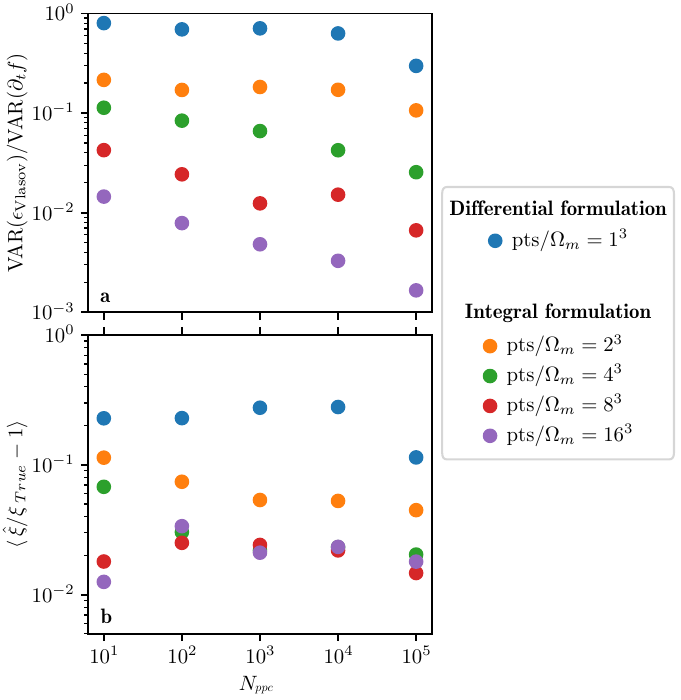}
\caption{\textbf{Impact of intrinsic data noise and size of integration volumes ($\Omega_m$) on the accuracy of the inferred PDE.} a) Deviation error from the true Vlasov equation $\epsilon_\mathrm{Vlasov}$ (defined as $\epsilon_\mathrm{Vlasov}(m) \equiv \int_{\Omega_m}~\left[ \partial_t f - v\partial_x f - q/m E\partial_v f\right]$) measured on two-stream data with varying numbers of particles per cell ($N_{ppc}$); the variance of this deviation error is normalized by the variance of $\int_{\Omega_m}\partial_t f$. b) Average relative error ($\langle \hat{\xi}/\xi_\mathrm{True} - 1\rangle$) of the inferred coefficients of the Vlasov equation using the sparse regression procedure, for varying $N_{ppc}$. In both plots, the different color points represent varying sizes of the integration volumes $\Omega_m$ used to sample each PDE term on the data. Each volume $\Omega_m$ is a cube of $n\Delta t \times n\Delta v \times n\Delta x$, with $n$ taking values between $1$ (single point volume) and $16$. Note that in varying the size of integration volumes (i.e. the number of points per volume, $\mathrm{pts}/\Omega_m$), we vary the number of integration volumes ($n_\Omega$) accordingly, so that the total number of points sampled from the data is fixed, with $N = \mathrm{pts}/\Omega_m \times n_\Omega = 512\mathrm{k}~\mathrm{pts}$.
}
\label{figS2}
\end{center}
\end{figure}

From the point of view of inferring PDE models from PIC simulation data, these fluctuations are a source of noise in the data and pose a challenge to the PDE inference procedure. Indeed, for typical $N_{ppc}$ used in PIC simulations, the data is too noisy for accurate and robust inference of PDEs using the point-wise sampling methods proposed in the original works on sparsity-based model identification (SINDy and PDE-Find). This is illustrated in Figure 1 (top row) of the main text, and the results obtained by this procedure for varying $N_{ppc}$ are summarized in Figure~\ref{figS2}. The blue points in Figure~\ref{figS2} (b) correspond to the point-wise evaluation strategy, and for $N_{ppc} = 10^1 - 10^4$ the mean error in the inferred coefficients of the Vlasov equation are significant, exceeding $20\%$. Even at the significantly higher computational cost of using $N_{ppc}=10^5$, the noise levels remain too high for accurate inference of the Vlasov equation coefficients, which are found with $10\%$ mean error. 

\begin{figure*}[t!]
\begin{center}
\includegraphics[width=1\textwidth]{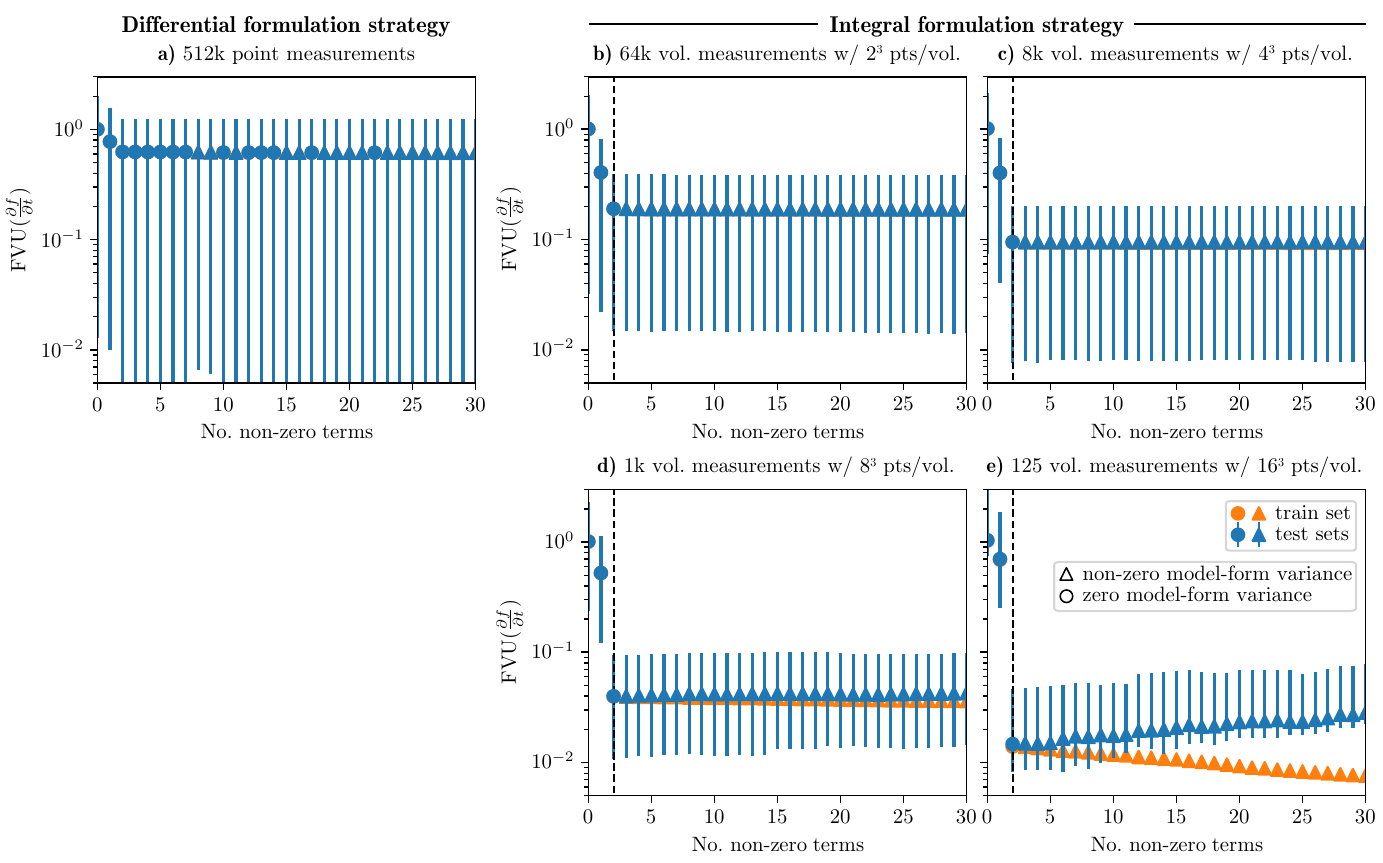}
\caption{\textbf{Impact of the size of the integration volumes on the accuracy/complexity curves.} Pareto analysis of model accuracy [$\mathrm{MSE(model)}/\mathrm{VAR}(\partial f/\partial t)$] versus the number of non-zero model terms for the inference of the Vlasov equation from two-stream data produced with $N_{pcc} = 10$ (corresponding to the highest level of discrete particle fluctuations explored here). Each panel corresponds to the accuracy/complexity curve obtained using different integration volume sizes $\Omega_m = n\Delta t \times n\Delta v \times n\Delta x$, with $n$ taking values: a) $1$ (single point volume, corresponding to the differential formulation), b) $2$, c) $4$, d) $8$ and e) $16$. As in Figure~\ref{figS2}, the total number of points used is kept fixed ($N=512\mathrm{k}~\mathrm{pts}$). The markers and error bars have the same meaning as in Figure~\ref{fig2}.
}
\label{figS3}
\end{center}
\end{figure*}

As discussed in the main text, we have circumvented the challenges posed by data noise by reformulating the problem of PDE discovery in its integral form. This is achieved by sampling the data in compact volumes ($\Omega_m$, instead of individual points) and evaluating the integral of each candidate PDE term over each of these volumes. %By varying the size of these integration volumes, we can effectively reduce the effects of noise on the evaluation of each PDE, and hence infer the underlying PDE more accurately and more robustly. 
Indeed, Figure~\ref{figS2} (a) shows that the deviation error of the true Vlasov equation evaluated on the data, $\varepsilon_\mathrm{Vlasov}$) relative to the ``physical signal'' (typical values of the time derivative of the distribution function, $\partial_t f$), decreases dramatically as we increase the size of the sampling volumes. This translates into a rapid reduction of the mean error in the inferred Vlasov coefficients with increasing size of the integration volumes, as seen in Figure~\ref{figS2} (b). For cubic volumes greater than $4\Delta t \times 4\Delta v \times4 \Delta x$, however, the mean coefficient error saturates at $\sim 1\%$. Further improvements are no longer observed because we have hit the level of the irreducible error in the data, which is associated with numerical discreteness and interpolation errors in the data generation procedure. More careful preparation of the data from the PIC simulations, so as to minimize interpolation errors associated with centering all quantities of interest in time an space, would lead to further improvements, but we leave this for future work.

It is also important to note the advantages of the integral formulation in improving the robustness of correct model identification in the presence of noisy data. This can be seen in Figure~\ref{figS3}, which shows the impact of varying the size of the integration volumes on the accuracy/complexity curves produced by the SR procedure on the two-stream data with the highest fluctuation levels (with $N_{ppc} = 10$). The large errors associated with the differential formulation lead to only a shallow increase in the model accuracy when an important dynamical term is removed [Figure~\ref{figS3}~(a)]. In addition, spurious extra terms that do not contribute to a significant reduction of the model error are systematically identified during 10-fold cross-validation (shown by the circle markers for models between $3-7$ terms). In contrast, the change in model accuracy when an important dynamical term is removed becomes progressively more pronounced as we increase the size of the integration volumes [Figure~\ref{figS3}~(b-e)]. This leads to a very clear signature of the correct underlying model. Note that the integral formulation also excludes the spurious extra terms that were previously being systematically identified in the differential formulation; cross-validation using the integral formulation shows high model-form variance for models with terms greater than $2$ terms (indicated by the triangle markers), meaning that different combinations of extra terms are identified in each cross-validation, and indicating that such extra terms are indeed spurious. For the largest volumes used in this example [$16\Delta t\times16\Delta v\times 16\Delta x$, Figure~\ref{figS3}~(e)], a clear divergence between the test and training errors is observed for models with more than $2$ terms, highlighting the loss of generalization for more complex models and further emphasizing the optimal accuracy/complexity trade-off at a model with $2$ terms; this divergence is obscured in the differential formulation case due its high sensitivity to data noise.

These results indicate that i) the performance of the integral formulation is weakly sensitive to the amount of discrete particle fluctuations in the data (controlled by $N_{ppc}$), ii) using the integral formulation on simulation data produced with few particles per cell yields more accurate inference of the PDE model coefficients than the use of the standard differential formulation on much cleaner data using many orders of magnitude more particles, and iii) the integral formulation significantly improves the robustness of the identification of the correct underlying PDE model from the accuracy/complexity curves.

\section{\label{app:sampling}Data sampling used for the inference of the underlying equations.} As noted in the main text, only a small fraction of the total data produced by each PIC simulation was used for the identification of the underlying equations via the SR procedure. The subset of the data in each example was obtained through \emph{uniform random sampling} in (phase) space and in time. A uniform and unbiased sampling strategy is important in general if no prior knowledge of the most relevant regions of the dynamics is available or assumed. Using this simple strategy, we show that we can recover the accurate form of the underlying equations. However, we note that this is not a very efficient strategy. Indeed, there are regions in the data where little or no dynamics is occurring -- for instance, during the near equilibrium phase of the plasma before the onset of the two-stream and Weibel instabilities, or in the quiet upstream plasma regions of the magnetized shock. The temporal and spatial derivatives of plasma quantities in these quiet regions are all close to zero and do not yield very useful information for the identification of the underlying PDEs. In scenarios where we are targeting specific physical processes and know in advance where they are dominant, we can focus our sampling on these regions allowing us to eliminate unnecessary data and making the application of these data-driven model discovery techniques more efficient.

\section{\label{app:thetadesign}Design of the library $\Theta$ for the inference of the multi-fluid and MHD equations.} The recovery of the multi-fluid equations in the main text was obtained through the nonlinear dynamics of the electron fluid undergoing the Weibel instability. The primary variables used to design the library $\Theta$ were naturally the moments of the electron distribution function ($n_e$, $n_e\langle \mathbf{v}\rangle_e$, $n_e\langle \mathbf{vv}\rangle_e$, $n_e\langle v^2\mathbf{v}\rangle_e$) and the self-consistent electric ($n_e\mathbf{E}$) and magnetic ($n_e\mathbf{B}$) force densities. We also considered the spatial gradients of all these variables up to second order (computed via centered finite differencing). The nonlinear candidate PDE terms were constructed by taking polynomial combinations of the primary variables and their gradients up to second order. 

For the recovery of the MHD equations, the library $\Theta$ was similarly constructed using the usual single-fluid variables ($\rho_m$, $\rho_m\langle \mathbf{v}\rangle$, $\rho_m \langle \mathbf{vv}\rangle$, $ \rho_m \langle v^2\mathbf{v}\rangle$, $\rho_c$, $\mathbf{J}$) based on the magnetized shock data; we also included the auxiliary variables $\mathbf{J'} \equiv \mathbf{J} - \rho_c \mathbf{v}$, $\mathbf{E'} \equiv \mathbf{E} + \mathbf{v}\times\mathbf{B}$, $\mathcal{P}\equiv \rho_m \langle \mathbf{vv}\rangle - \rho_m\langle \mathbf{v}\rangle \langle \mathbf{v}\rangle$ and $p\equiv (\sum_i \mathcal{P}_{ii})/3$ in $\Theta$ for inference of the MHD energy equation.

\begin{figure*}[t!]
\begin{center}
\includegraphics[width=1\textwidth]{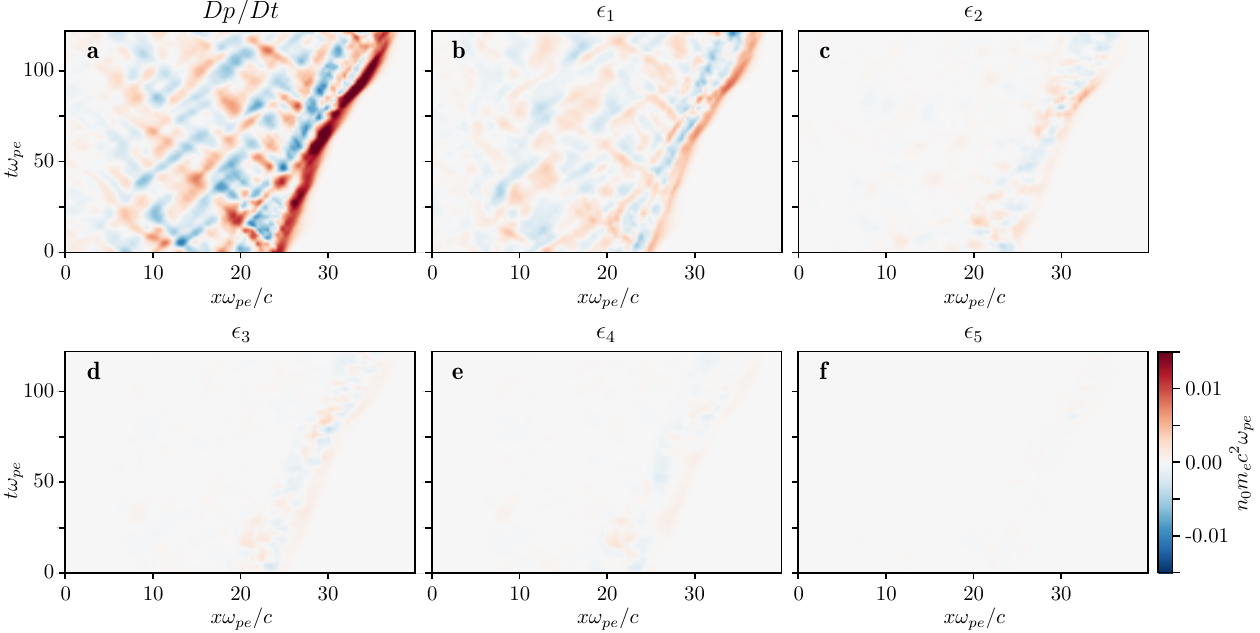}
\caption{\textbf{Spatiotemporal error distribution of the hierarchy of Pareto-optimal models for the MHD energy density equation.} a) Total time derivative of the scalar pressure field ($Dp/Dt \equiv \partial_t p + \langle v_x\rangle \partial_x p$) evaluated on the magnetized shock data. Figures b)-f) reveal the spatiotemporal error distribution of the inferred Pareto-optimal models presented in Figure~4 of the main manuscript. The error distribution of a model with $M$ terms is denoted by $\epsilon_M \equiv Dp/Dt - \Theta \hat{\xi}_M$, where $\hat{\xi}_M$ contains the $M$-term model coefficients inferred by the sparse regression procedure.
}
\label{figS4}
\end{center}
\end{figure*}

\section{\label{app:spatiotemporal}Analysis of the spatiotemporal error distribution of inferred PDE models}
The sparse regression procedure aims to find the most parsimonious model that minimizes the mean squared error (MSE) on the sampled data points (or volumes $\Omega_m$ in case of the integral formulation). However, once a model has been identified, assessing its ``quality'' (its ability to describe the physics/dynamics in the data) cannot readily be determined from its MSE alone. Indeed, it is important to discern between two sources of error: i) irreducible error, associated with contributions from data noise (and intrinsic numerical discretization and interpolation errors in simulation data), and ii) model bias error, associated with potential missing terms (missing physics) in the library $\Theta$. Unless a detailed understanding of the irreducible error in the data is available, a useful diagnostic to discern between these two sources of error is the analysis of the spatial and temporal error distribution of the model on the data.

This is illustrated in Figure~\ref{figS4} for the hierarchy of Pareto-optimal models inferred for the single-fluid energy density (pressure) equation from the magnetized shock data (and which is presented in Figure~4 of the main text). The total time derivative of the pressure field ($Dp/Dt$) is shown in Figure~\ref{figS4} (a), revealing a quiet upstream ahead of the shock, a rapid increase in the pressure field at the shock front, and wave structures in the downstream plasma behind the shock. The discrepancy errors $\epsilon_M$ for models of increasing complexity (increasing number of terms, $M$) are presented in Figures~\ref{figS4} (b-f). While the overall MSE (or FVU, fraction of variance unexplained) of each model expectedly decreases with increasing model complexity, the spatiotemporal distribution of the model error offers valuable insights into the physics captured by each model and the impact of its approximations. Figure~\ref{figS4} (b), for instance, shows the error distribution of the simplest $1$-term model (corresponding to the adiabatic closure model), which neglects heat flux, pressure anisotropy, and Joule heating, capturing only compressional heating. The coherent structures seen in the error distribution, which closely correlate with the structures seen in $Dp/Dt$, indicate that this simplified model provides an overall crude approximation to the dynamics in the data. The more refined $2$-term model that includes finite heat flux physics, leads to a much improved description of the plasma dynamics downstream of the shock, leaving only the shock transition itself poorly described [Figure~\ref{figS4} (c)]. This indicates that there remains missing physics in the model to correctly describe the shock transition region. Indeed, it is well known that this region is characterized by high pressure anisotropy and Joule heating, and the inclusion of these effects in the more complex models show progressively improved approximations of the plasma dynamics at the shock transition [Figures~\ref{figS4} (d-f)].

Therefore, the spatiotemporal distribution of the model error on the data not only elucidates where and when different dynamical terms are important, but can also serve as an indicator of missing physics (informing on incompleteness of the library of candidate terms $\Theta$ used in the regression). This can be extremely useful when applying this methodology to infer models of complex and poorly understood plasma phenomena. By pinpointing where and when the model errors are most significant, the spatiotemporal error distribution provides important clues as to what the missing physics is, and serves as a guide on how to expand and refine the library $\Theta$.

\nocite{*}

\bibliography{scibib}% Produces the bibliography via BibTeX.

\end{document}